\documentclass{camera}
\usepackage{graphicx}  

\begin{document}

%
\title{Ultra high energy cosmic rays and neutrinos after Auger}

%
\author{Todor Stanev}

%
\organization{Bartol Research Institute, Department of physics and Astronomy,
 University of Delaware, Newark, DE 19716, U.S.A.}

\maketitle

\begin{abstract}
 We discuss the main results that were recently published by the 
 Auger Collaboration and their impact on our knowledge of the
 ultra high energy cosmic rays and neutrinos.
\end{abstract}

%

\section{Introduction}

 We have discussed the problems related to the ultra high 
 energy cosmic rays (UHECR) quite many times in recent years
 after the results of the high resolution Fly's Eye (HiRes) 
 started trickling out several years ago. The cosmic ray
 energy spectrum extracted from the the HiRes observations
 indicated that the spectrum suffers a strong decrease at
 energies above 5$\times$10$^{19}$ eV~\cite{HiRes_sp}, 
 consistent with the GZK cutoff~\cite{GZK} that is due
 to UHECR interactions with the microwave background. There
 was thus a strong disagreement with the energy spectrum
 derived from the previous largest exposure cosmic ray
 air shower array, Agasa~\cite{Agasa} that seemed to 
 continue with the same slope above 10$^{20}$ eV.

 HiRes and Agasa measure the cosmic ray events in different
 ways. HiRes is a fluorescent detector that follows the shower
 longitudinal development by detecting the fluorescent light
 emitted by the Nitrogen atoms in the air ionized by the shower
 charged particles. The primary cosmic ray energy is obtained
 from integration of the shower longitudinal profile 
 normalized by the average electron energy loss and the 
 fluorescent efficiency of the air. Agasa, on the other hand,
 is a classical air shower array that records the particle
 density at the observation level and uses Monte Carlo 
 calculations to relate it to the primary cosmic ray energy.
 The dependence of this method on the hadronic interaction 
 model used in the simulations is considered stronger in this
 approach.

 In 2007 the first results of the Pierre Auger Observatory (PAO) 
 in Argentina were published. This experiment became the hope 
 for better understanding of UHECR several years ago when its
 construction started. It is a hybrid air shower experiment that
 applies both detection methods. PAO consists of an air shower 
 array of area 3,000 km$^2$ where the individual detectors are
 10 m$^2$ water tanks placed at distances of 1.5 km in a 
 hexagonal pattern (SD). 
 The shower array is surrounded by 24 
 fluorescent telescopes (FD) located in four groups of six. The idea
 is that during cloudless moonless nights when the fluorescent 
 detectors can work the UHECR showers will be observed 
 simultaneously by both types of detectors and this {\it hybrid}
 observation will unveil better the shower properties. Here is a brief
 summary of the Pierre Auger Observatory results. 

 The energy spectrum derived from PAO data~\cite{auger_sp} shows a
 feature consistent with the GZK cutoff similar, but not identical
 to that of HiRes - see Fig.~\ref{AugerHR}. The spectrum is based 
 on the much higher surface detector statistics normalized to the
 energy estimates of the fluorescent detector in hybrid events.
 The problem is that if the energy was estimated only from the
 SD data (particle density at 1,000 meters from the shower core - 
 $S_{1000}$) the energy scale would be 25\% higer~\cite{auger_re}.
 Formally this is not a big problem because the uncertainty of
 the energy estimate is 22\%.

 Another problem appears when PAO attempts to estimate the type
 of the primary cosmic ray nuclei. When it is estimated from the
 depth of the shower maximum measured by the fluorescent 
 detector $X_{max}$ the average primary mass appears to be
 close to He~\cite{auger_mu} . If the surface detector density
 $S_{1000}$ is used to estimate the number of shower muons~\cite{auger_re}
 the average primary nuclei appear to be much heavier.
 The preliminary conclusion from these two disagreements is
 that showers are absorbed in the atmosphere slower than models
 predict. 

 The Auger Collaboration succeeded, however, to eliminate a class
 of UHECR models that predicted that the primary particles 
 are gamma rays. From studies of $X_{max}$ and other shower
 properties the fraction of gamma rays in the cosmic rays above
 10$^{19}$ eV is limited to not more than 2\%~\cite{auger_g}.
 The much smaller statistics at higher energy does not allow 
 for such strong limits, but at least the bulk of UHECR, which
 we consider extra galactic, consists of nuclei.

 The Auger result that impressed the most not only the
 astrophysical community, but also the general public, was the
 published correlation of their highest energy events with
 active galactic nuclei~\cite{Auger1,Auger2}.
 Twenty of the 27 events of energy above 5.7$\times$10$^{19}$ eV
 (57 EeV) come within 3.1$^o$ of the position of AGN at 
 redshift smaller than $\leq$0.017 (distance of 71 Mpc) while one
 expects only 7 in case of isotropic source distribution.
 The 3.1$^o$ circle around the event direction (that was 
 obtained from a scan in energy, angle and distance) is understood
 as scattering in the galactic magnetic field plus about 1$^o$ 
 Auger angular resolution. Accounting for the number of scans 
 the chance probability is well below 1\%. If the events that
 come from galactic latitude less than 10$^o$ (where catalogs
 are not full and cosmic ray may scatter more) are excluded
 then 19 out of 21 events are closer than 3.1$^o$ from nearby AGN.

 Since the Auger collaboration published the arrival directions
 and energies of these 27 events in Ref.~\cite{Auger2} this
 announcement was quickly followed by a number of other 
 analyses and questions, like:\\
 \hspace*{5mm} $\bullet$ why only 71 Mpc when $>$60 EeV cosmic rays
 should come from  distances up to 200 Mpc ?\\
 \hspace*{5mm} $\bullet$ why there are no events from the direction
 of the Virgo cluster, that contains powerful AGN ?\\
 \hspace*{5mm} $\bullet$ what does the concentration of events from
 directions close to Cen A means ?\\
 Many of these questions were also asked by the Auger
 Collaboration~\cite{Auger2}.

\section{UHECR energy spectrum}

\subsection{Formation of the cosmic ray spectrum in propagation}
 The energy spectrum at energies above 10$^{18}$ eV depends mostly
 on two parameters: the acceleration spectrum at the sources
 and the source distance distribution, as the spectrum
 detected at Earth evolves in propagation in the universal 
 photon fields. Most of the propagation calculations used for
 the determination of the acceleration spectrum have been done
 in two simplifying assumptions:\\
 $\bullet$ All sources have identical acceleration spectra, and\\
 $\bullet$ Sources are isotropically distributed in the Universe.

\begin{figure}[thb]
\centerline{\includegraphics[width=10truecm]{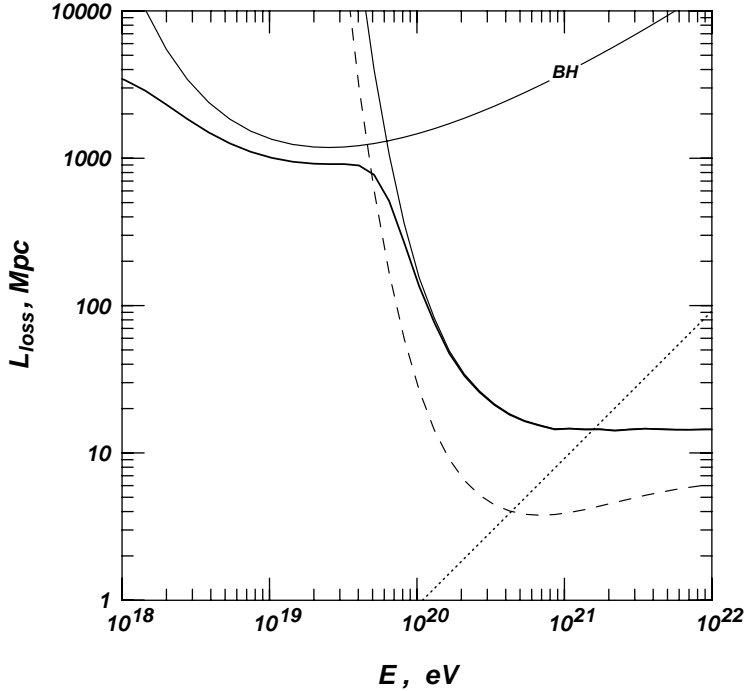}}
\caption{ The dashed line shows the photoproduction interaction
 length and thick solid one - the total energy loss length 
 including the pair production process, labeled as BH. The neutron
 decay length is plotted with a dotted line.}
\label{xloss} 
\end{figure}

 Since the energy of the background photons is much lower than 
 1 eV the center of mass energy for $p\gamma$ interactions is
 low, at the threshold for photoproduction interactions. 
 The cross section of the $p\gamma$ photoproduction interactions
 is very well known from
 accelerator measurements - see Fig.~3 in Ref.~\cite{muckeetal00}
 as well as the fractional proton energy loss per interaction.
 Another energy loss process for protons is the $e^+e^-$
 Bethe-Heitler  pair
 production~\cite{hillas75,BerGri} where the proton energy loss 
 peaks at about 10$^{19}$ eV and the adiabatic energy loss due
 to the expansion of the Universe. The energy loss length for
 protons is shown in Fig.~\ref{xloss}.

 Nuclei heavier than H in addition to photoproduction and
 pair production (scaled by $Z^2$ and shifted up in total 
 energy by $A$) suffer from a different energy loss - 
 photo disintegration~\cite{Stecker69}. There are two main processes - 
 one nucleon loss at the giant nuclear resonance and two nucleon 
 loss at higher photon energy in the nucleus frame. One has also
 to account for the decay of unstable nuclei generated this way.

 If UHECR are protons there are two general solutions for the
 cosmic ray acceleration spectrum - relatively steep 
 acceleration spectrum $E^{-(\gamma + 1)}$ with no cosmological
 evolution of the 
 sources (a l\'{a} Berezinsky~\cite{Berezinskyetal}) and 
 a much flatter ($\gamma$ = 1.2-1.3) with a strong cosmological 
 evolution such as that of the star forming regions (SFR).
 The cosmological evolution is not important for the highest
 energy cosmic rays that can not come to us from redshifts 
 higher than 0.05 but does affect lower energy ones. Note that
 the first (`dip') model does not need any galactic cosmic rays 
 above about 10$^{18}$ eV. 
 Mixed composition models~\cite{Allardetal} generally do not require
 any cosmological evolution of the UHECR sources and are consistent
 with acceleration spectra with $\gamma$ = 1.2-1.3.
  
 \subsection{Experimental data}
\begin{figure}
\centerline{\includegraphics[width=5truein]{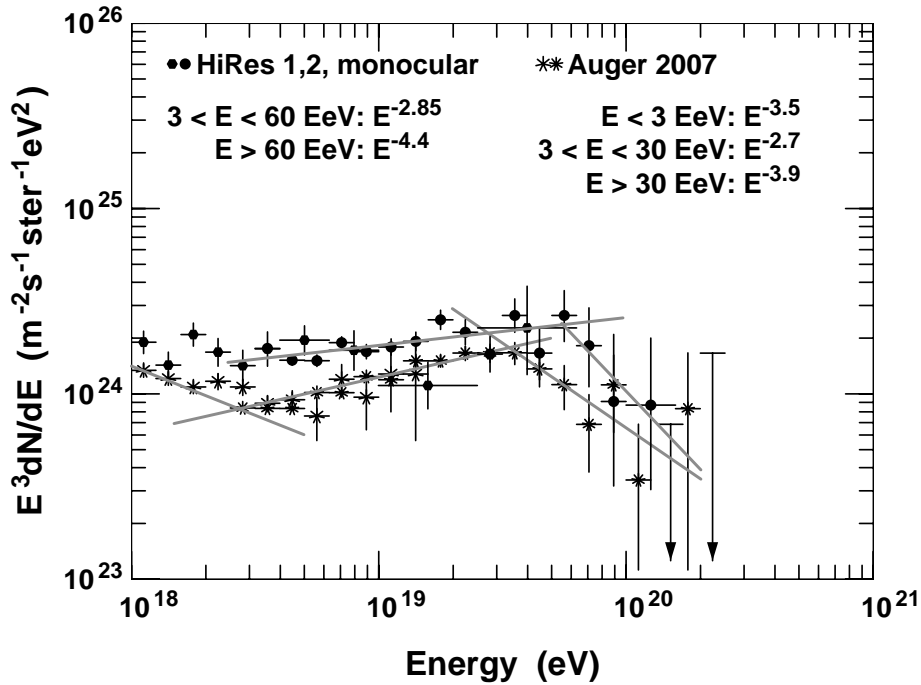}}
\caption{Cosmic ray energy spectra derived from data by the
 Auger and HiRes Collaborations. The two types of data points for
 Auger give the SD and hybrid data sets. The monocular spectra of
 HiRes 1 and 2 are shown for HiRes.}
\label{AugerHR}
\end{figure}

 Figure~\ref{AugerHR} compares the cosmic ray energy spectra 
 derived by the Auger and the HiRes Collaborations. The 
 lines drawn through the data points are rough fits of the 
 spectra in limited energy ranges. Note these are not identical
 to the fits made by the Auger collaboration~\cite{auger_ty}.
 Apart from the overall normalization of the two spectra they
 show a small difference in shape that is still statistically not
 very significant but currently favor different astrophysical 
 models. The HiRes spectrum~\cite{hires_mono} is fully consistent with
 the model of Berezinsky et al that involves a steep acceleration spectrum
 ($\gamma$ = 1.7). The `dip' at about 3$\times$10$^{18}$ eV is formed
 at the edge of  pair production and adiabatic energy loss. 
 There are several good fits to the Auger energy spectrum performed by
 the collaboration itself~\cite{auger_ty}: a steep proton acceleration 
 spectrum with $\gamma$ = 1.55 without cosmological evolution, a flatter
 one with $\gamma$ = 1.3 and strong $(1 + z)^5$ cosmological evolution,
 and a mixed composition spectrum with $\gamma$ = 1.2  Some of these
 fits require the galactic cosmic ray spectrum to extend above
 10$^{18.5}$ eV. It is worth noting that if the Auger energy scale 
 were 25\% higher, the spectrum would be much closer to the HiRes
 one in both overall normalization and in the position of the
 GZK feature and that they predict different chemical compositions
 of UHECR.
  
 \section{Secondary particle fluxes from cosmic ray propagation}

 The energy that UHECR lose in propagation end up in fluxes of 
 gamma rays and neutrinos from $\pi^0$ and $\pi^\pm$ decays.
 Gamma rays have even shorter energy loss length than nuclei
 and develop pair production/inverse Compton cascades where 
 synchrotron radiation may play important role. In the presence 
 of noticeable (1 nG) extragalactic field 10$^{18}$ eV electrons 
 quickly lose energy to GeV synchrotron photons.

 Neutrinos, however, have very low interaction cross section 
 and arrive at Earth with just redshift energy loss from
 almost any distance. For this reason the cosmological 
 evolution of the cosmic ray sources are very important input
 in the calculation of these {\em cosmogenic}~\cite{BerZat,Steck}
 neutrinos. 

 Two among the Auger results presented above may have an
 important connection to the flux of cosmogenic neutrinos -
 the correlation of the highest Auger events with nearby 
 AGN and the $(1+z)^3$ cosmological evolution evolution  
 required by one of the spectrum fits with relatively 
 flat ($\gamma$ = 1.3) proton models.

 \subsection{Cosmological evolution of AGN} 
 
 The cosmological evolution of AGN has been studied by their 
 X-ray emission as a function of the redshift. Most of the 
 X-ray satellite measurements were sensitive to relatively
 low energies and the worlds richest data set is 
 in the 0.5 to 2 KeV range. Anyway it is very difficult to 
 know what the best proxy is for the evolution of the 
 cosmic ray luminosity, but the available astronomical data
 are either in infrared (for the SFR) or in low energy X-rays.

 The AGN observations of the Rosat satellite were analyzed
 in Ref.~\cite{FSAD98} and the derived cosmological evolution
 is $(1+z)^5$. A later analysis~\cite{HMS05} used much
 larger statistics and was able to divide the observed AGN
 in four luminosity groups, each encompassing one order of 
 magnitude above 10$^{42}$ erg/Mpc$^3$. It turns out, according
 to this analysis, that the highest X-ray luminosity AGN 
 also have the highest luminosity evolution: $(1+z)^7.1$
 up to $z_{max}$ = 1.7. Such cosmological evolution would 
 indeed generated very high fluxes of cosmogenic neutrinos.

\begin{figure}[thb] 
\centerline{\includegraphics[width=4.5 in]{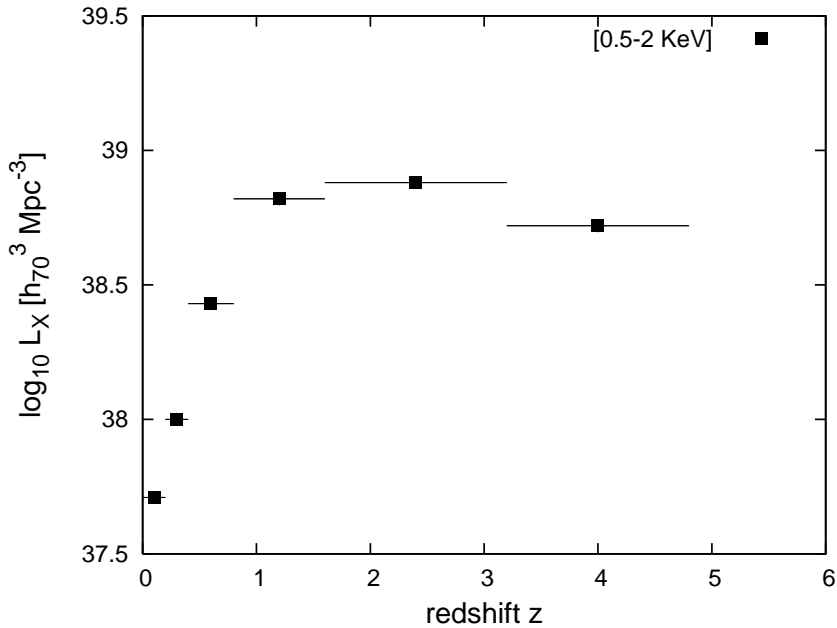}}
\caption{Total X-ray AGN luminosity as a function of redshift.
 Values are integral from the luminosity table of Ref.~\protect
 \cite{HMS05}.
\label{table3z}
}
\end{figure}

 There are not, however, many such AGN and it is not very likely 
 that they are the sources of UHECR that have to come (according
 to the analysis in Ref.~\cite{Auger2}) from many nearby sources.
 It maybe better to use for this purpose the cosmological evolution
 of the AGN X-ray luminosity that could be obtained by summation
 from the data published in Ref.~\cite{HMS05} and is shown 
 in Fig.~\ref{table3z}.
Up to redshifts exceeding 1 the total AGN luminosity has 
$(1+z)^5$ cosmological evolution, significantly stronger
than the evolution of star forming regions. 

\subsection{Fluxes of cosmogenic neutrinos}

 In addition to the cosmological evolution of the
 cosmic ray sources the flux of cosmogenic neutrinos depends
 on a number of  parameters:
 \begin{itemize}
 \item The average acceleration slope of the highest energy
 cosmic rays (UHECR)
 \item The maximum acceleration energy $E_{max}$.
 \item The distribution of the cosmic ray sources in the Universe.
 This is important to estimate the average UHECR emissivity
 since we only observe UHECR generated in our cosmological 
 neighborhood, i.e. $z <$ 0.40. Auger analysis requires
 source density higher than 3.5$\times$10$^{-5}$ Mpc$^{-3}$. 
 (Number sources within 75 Mpc > 61)
 \end{itemize}

 We will use the following parameters in calculation of the cosmogenic
 neutrino flux:\\
 UHECR emissivity of 2.25$\times$10$^{44}$ erg/Mpc$^3$, i.e. one half of
 the value derived by Waxman and used in an earlier
 calculation~\cite{ESS}\\
 acceleration spectrum with $\gamma$ = 1.3 with 
 an exponential cutoff at 10$^{21.5}$ eV\\
 cosmological evolution $(1+z)^5$ up to $z_{max}$ = 1.7, then
 flat to $z$ = 2.7 with exponential decline after that\\
 cosmological model with $\Omega_\Lambda$ = 0.73\\
 and new neutrino yields calculated with the same cosmological model
 for interactions on the microwave background

\begin{figure}[thb]
\centerline{\includegraphics[width=4.5 truein]{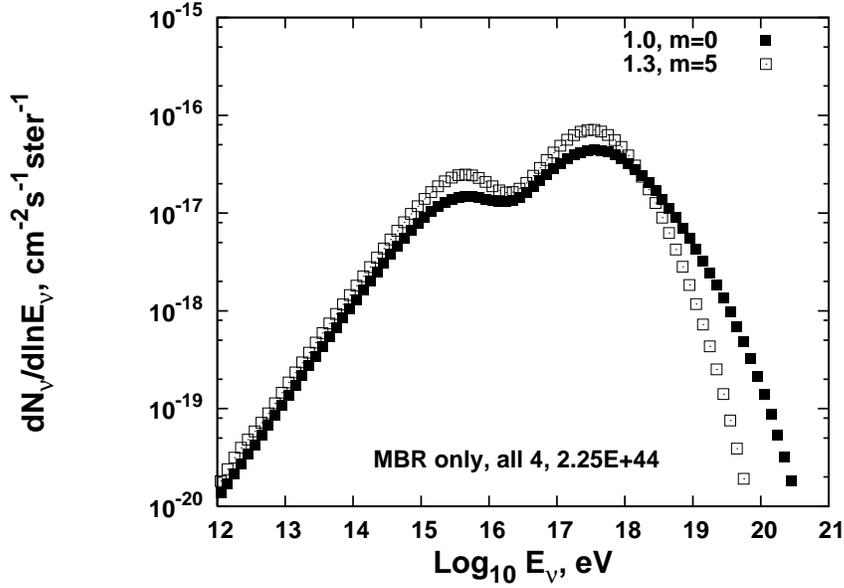}}
\caption{ Comparison of the flux of cosmogenic neutrinos to
 that of Ref.~\protect\cite{ESS}}
\label{all4} %
\end{figure}

 Figure~\ref{all4} compares  the spectrum of all four neutrino 
 flavors calculated in this way to that of the previous 
 calculation based on Waxman's parameters~\cite{Waxman95}.
 Note that the higher energy
 peak consists mostly of $\nu_\mu$, $\bar{\nu}_\mu$ and $\nu_e$ while 
 the lower energy one represents $\bar{\nu}_e$ from neutron decay.
 The contribution of electron anti neutrinos to the higher energy
 peak comes from neutron interactions and is small as most neutrons decay.
 Although the fluxes are not very distant from each other there are
 several significant differences:\\
 $\bullet$ The Auger inspired flux cuts off at lower energy because of its
 steeper acceleration spectrum in spite of the identical maximum energy.
 If the maximum energy were lower the total normalization of the
 flux would be lower and the cutoff would be scaled down.\\
 $\bullet$ The peaks in the neutrino energy distribution are slightly
 higher but at somewhat lower energy. Both effects can be related 
 to the higher contribution of high redshifts.

 It is difficult to estimate correctly the detectable event rates
 from these two distributions, but the Auger inspired fit is 
 certainly much less efficient in production of very high energy
 neutrinos that could be detected by experiments such as ANITA
 and generally all experiments based on radio detection of
 neutrino induced showers.

 A full calculation of the cosmogenic neutrino fluxes should also 
 include neutrinos from interactions in the infrared/optical 
 background. In such a calculation the relatively flat acceleration
 spectrum would not make a huge difference although the infrared
 background contribution would be stronger than the one for
 $\gamma$ = 1 spectrum. The strong cosmological evolution of the 
 sources would also not be very important since the infrared
 background evolves not as fast as the microwave one.

 \subsection{Effects of the cosmic ray composition}

 Models of UHE cosmic rays accelerated in a mixed cosmic ray
 composition generate cosmogenic neutrino fluxes with
 a different shape. Since the energy measured by air showers
 is the total energy per nucleus such models generate 
 smaller high energy peaks and much higher lower energy one
 from neutron decays after disintegration. The photoproduction
 threshold is proportional to the nuclear mass, $A$ times higher
 then 3$\times$10$^{19}$ which is the proton threshold for 
 interactions in MBR. He nuclei would then start interacting
 at energy higher than 10$^{20}$ eV and iron nuclei - above
 10$^{21}$ eV. Although the maximum energy achieved in
 acceleration depends on the nuclear charge $Z$ such nuclei
 may not exist and the mixed composition models tested by
 Auger only extend to 10$^{20}$ (10$^{21}$) eV, i.e. the
 energy per nucleon does not reach the He (Fe) photoproduction
 threshold.

 The first approximation for the flux of cosmogenic neutrinos 
 generated by mixed composition models is that only their
 proton component undergoes photoproduction interactions. 
 This, however, is not well defined either because the
 cosmic ray composition continuously changes in propagation -
 it becomes lighter as nuclei disintegrate. A good estimate
 can only be obtained by a direct Monte Carlo simulation 
 assuming a particular cosmic ray composition at the sources.
 Assuming a composition introduces some more free parameters
 in the calculation. It is important to note that the lower
 energy $\bar{\nu}_e$ peak is below the Glashow resonance
 energy and would not give a high neutrino interaction rate.

\section{Conclusions}

 The first results of the Auger Collaboration did improve
 our knowledge of the ultra high energy cosmic rays. They
 are fully consistent with a GZK feature in the cosmic ray
 spectrum as suggested previously by the HiRes Collaboration.
 Being a hybrid experiment Auger demonstrated that the energy
 estimate by fluorescent detectors and surface air shower array
 is different, as previously suggested by the Agasa and HiRes
 spectra. Now the UHE community is searching for the reasons
 for this disagreement that may hide in the currently used 
 hadronic interaction models.

 Auger started the direct search for the sources of UHECR by 
 studies of the correlation of their highest energy events with
 active galactic nuclei. The published correlation inspired 
 many other different analyses of correlation with powerful
 astrophysical objects. This work is in progress now and the
 UHECR sources will be identified eventually when the Auger 
 statistics grows - it will double at the end of August 2008.

 Since the current UHECR energy spectrum can be fit with 
 several different models it is not obvious that we know better
 the expected flux of cosmogenic neutrinos. This may only happen
 when an unique fit of the spectrum is achieved.\\[3truemm]
 {\bf Acknowledgments} This talk is based on collaboration with
 many colleagues including Peter Biermann, Daniel DeMarco,
 Thomas Gaisser, J\"{o}rg Rachen, \& David Seckel. The author was
 supported in part by NASA APT grant NNG04GK86G.


\vspace*{3truecm}
\noindent
{\bf DISCUSSION}\\[5truemm]
{\bf DANIELE FARGION} In your last paper few weeks ago the
 main conclusive sentence was: Following our 1995 paper one
 could think the Auger events may reach us from Cen A. 
 I could not find this prediction in that paper.\\[2truemm]
{\bf TODOR STANEV}\\[2truemm] I should not have used a similar
 statement in the conclusions since this option was not really
 discussed in the paper. What I had in mind and tried to lead
 to is that with reasonable magnetic fields (order of 1 $\mu$G) 
 in the Cen A lobe one can imagine UHECR accelerated at that 
 object to bend at significant angles and appear to the 
 observer as coming from nearby sources.\\[2truemm]
{\bf SERGIO COLAFRANCESCO} It seems to me that you are assuming a single
 population of AGN in your calculations and not accounting of their 
 astrophysical differences. What are the implications of such an
 assumption  on your results.\\[2truemm]
{\bf TODOR STANEV} You are absolutely correct, Sergio. All
 current estimates are done in the assumption that all
 sources have  identical parameters. There are only a few
 attempts to account for differences choosing a certain
 type of source and a luminosity and maximum energy for
 each source (see P.L.~Biermann and students work).
 Unfortunately in such analyses one has to rely 
 on simple theoretical calculations of these important
 parameters. One can easily fit the UHECR spectrum in
 such models as they introduce more `free' parameters.

%
\end{document}